\documentclass[a4paper]{jpconf} 
\usepackage{graphicx}
\usepackage{amssymb,amsmath} 

\def\pr{\prime}
\def\be{\begin{equation}}
\def\lan{\left\langle}
\def\ran{\right\rangle}
\def\ee{\end{equation}}
\def\barr{\begin{array}}
\def\earr{\end{array}}

\def\nn8{\\}
\def\l{\left}
\def\r{\right}
\def\dis{\displaystyle}
\def\ed{\end{document}}

\def\cod{{\cal O}^\dagger}
\def\co{{\cal O}}

\def\wtM{{\widetilde {M}}}

\begin{document}
\title{Group Theory for Embedded Random Matrix Ensembles}

\author{V.K.B. Kota}

\address{Physical Research Laboratory, Ahmedabad 380 009, India}

\ead{vkbkota@prl.res.in}

\begin{abstract}

Embedded random matrix ensembles are generic models for describing statistical
properties of finite isolated quantum many-particle systems. For the simplest
spinless fermion (or boson) systems with say $m$ fermions (or bosons) in $N$
single particle states and interacting with say $k$-body interactions, we have
EGUE($k$) [embedded GUE of $k$-body interactions)  with GUE embedding and the
embedding algebra is $U(N)$.  In this paper, using EGUE($k$) representation for
a Hamiltonian  that is $k$-body and an independent EGUE($t$) representation for a
transition  operator that is $t$-body and employing the embedding $U(N)$ 
algebra, finite-$N$ formulas for moments  up to order four are derived, for the
first time, for the transition strength  densities (transition strengths
multiplied by the density of states at the initial and final energies).  In the
asymptotic limit, these formulas reduce to those derived for the EGOE  version
and establish that in general bivariate transition strength  densities take
bivariate Gaussian form for isolated finite quantum systems. Extension of these
results for other types of transition operators and EGUE ensembles with further
symmetries are discussed.

\end{abstract}

\section{Introduction}

Wigner introduced random matrix theory (RMT) in physics in 1955 primarily to
understand statistical properties of neutron resonances in heavy nuclei
\cite{Po-65,Widen}.  Depending on the global symmetry properties of the
Hamiltonian of a quantum system, namely rotational symmetry and time-reversal
symmetry, we have Dyson's tripartite classification of random matrices giving
the classical random matrix ensembles, the Gaussian orthogonal (GOE), unitary
(GUE) and symplectic (GSE) ensembles. In the last three decades, RMT has found
applications not only in all branches of quantum physics but also in many other
disciplines such as Econophysics, Wireless communication, information theory,
multivariate statistics, number theory, neural and biological networks and so on
\cite{rmt1,rmt2,rmt3,rmt4}.  However, in  the context of isolated finite
many-particle quantum systems, classical random matrix ensembles are too
unspecific to account for important features of the physical system at hand. One
refinement which retains the basic stochastic  approach but allows for such
features consists in the use of embedded random matrix ensembles. 

Finite quantum systems such as nuclei, atoms, quantum dots,  small metallic
grains, interacting spin systems modeling quantum computing core and ultra-cold
atoms, share one common property - their constituents predominantly interact
via  two-particle interactions. Therefore, it is more appropriate to represent
an isolated  finite interacting quantum system, say with $m$ particles (fermions
or bosons) in $N$ single particle (sp) states by random matrix models generated by
random $k$-body (note that $k < m$ and most often we have $k=2$)  interactions
and propagate the information in the interaction to many particle spaces. Then
we have random matrix ensembles in $m$-particle spaces - these ensembles are
defined by representing the $k$-particle Hamiltonian ($H$) by GOE/GUE/GSE and
then the $m$ particle $H$ matrix is   generated by the $m$-particle Hilbert
space geometry. The key element here is  the recognition that there is a Lie
algebra that transports the information in the two-particle spaces to
many-particle spaces.  As a GOE/GUE/GSE random matrix ensemble in two-particle
spaces is embedded  in the $m$-particle $H$ matrix, these ensembles are  more
generically called embedded ensembles (EE). 

Embedded ensembles are proved to be rich in their content and wide in their 
scope. A book giving detailed discussion of the various properties and
applications of a wide  variety of embedded matrix ensembles is now available
\cite{kota}. Significantly, the study of embedded random matrix ensembles is
still  developing. Partly this is due to the fact that mathematical tractability
of these ensembles is still a problem and this is the topic of the present
paper. A general formulation for deriving analytical results is to use the
Wigner-Racah algebra of the embedding Lie algebra \cite{kota}. The focus in the
present paper is on transition strengths. Note that transition strengths probe
the structure of the eigenfunctions of a quantum many-body system.

Given a transition operator $\co$ acting on the $m$ particle eigenstates
$\l|\l.E\ran\r.$ of $H$ will give transition matrix elements $\l|\lan E_f \mid
\co \mid E_i\ran\r|^2$ and the corresponding transition strength density
(this will take into account degeneracies in the eigenvalues) is,
\be
I_\co(E_i,E_f) = I(E_f)\,\l|\lan E_f \mid \co \mid E_i\ran\r|^2\,I(E_i) \;.
\label{strn-eq1}
\ee
In Eq. (\ref{strn-eq1}), $I(E)$ are state densities normalized to the dimension
of the $m$ particle spaces. Note that $E_i$ and $E_f$ belong to the same $m$
particle system or different systems depending on the nature of the transition
operator $\co$. In the discussion ahead, we will consider both situations.
Random matrix theory has been used in the past to derive the form of $I(E)$, 
the state densities. In particular, exact (finite $N$) formulas for lower order 
moments $\lan H^p\ran$, $p=2$ and $4$ of $I(E)$ are derived both for EGOE and
EGUE ensembles using group theoretical methods directly \cite{Ko-05} or
indirectly \cite{Be-01}. Going beyond the eigenvalue densities, here we will
apply group theoretical methods and derive for the first time some exact
formulas for the  lower order moments of the transition strength densities
$I_\co(E_i,E_f) $. Using these, the general form of transition strength
densities is deduced.  We will restrict to fermion systems and at the end
discuss the extension to boson systems and to other situations. Let us add that
some results valid in the asymptotic ($N \rightarrow  \infty$) limit for both
the state densities and transition strength densities  are available in
literature \cite{Mo-75,FKPT,Ma-arx,Muller}. 

In general, the Hamiltonian may have many symmetries with the fermions (or
bosons) carrying other degrees of freedom such as spin, orbital angular
momentum, isospin and so on. Also we may have in the system different types of
fermions (or bosons) and for example in atomic nuclei we have protons and
neutrons. In addition, a transition operator may preserve particle number and
other quantum numbers or it may change them. Off all these various situations,
here we have considered three different systems: (i) a system of $m$ spinless
fermions and a transition operator that preserves  the particle number;  (ii) a
transition operator that  removes say $k_0$ number of particles from the $m$
fermion system;  (iii) a system with two types of spinless fermions with the
transition operator changing $k_0$ number of particles of  one type to $k_0$
number of other particles as in nuclear beta and double beta decay. In all
these we will restrict to EGUE. Now we will give a preview.

Section 2 gives some basic results for EGUE($k$) for spinless fermion systems as
derived in \cite{Ko-05} and some of their extensions. Using these results,
formulas for the lower order bivariate moments of the transition strength
densities for the situation (i) above are derived and they are presented in
detail in Section 3. Using these, results in the asymptotic limit are derived and
they are presented in Section 4. Some basic results for the situations (ii) and
(iii) above are given in Section 5. Finally, in Section 6 briefly discussed are
some open group theoretical problems in the embedded ensembles theory for 
transition strength densities.

\section{Basic EGUE($k$) results for a spinless fermion system}

Let us consider $m$ spinless fermions in $N$ degenerate sp states with the 
Hamiltonian $\hat{H}$ a $k$-body operator,
\be
\hat{H} = \sum_{i,j} V_{ij}(k)\;A^\dagger_i(k)\,A_j(k)\;,\;\;\;V_{ij}(k)
=\lan k,i \mid \hat{H} \mid k,j\ran\;.
\label{strn-eq2}
\ee
Here $A^\dagger_i(k)$ is a $k$ particle (normalized) creation operator and
$A_i(k)$ is the corresponding annihilation operator (a hermitian conjugate). 
Also, $i$ and $j$ are $k$-particle indices. Note that the $k$ and $m$ particle
space dimensions are $\binom{N}{k}$ and $\binom{N}{m}$ respectively. We will
consider $\hat{H}$ to be EGUE($k$) in $m$-particle spaces. Then $V_{ij}$ form a
GUE with $V$ matrix being Hermitian. The real and imaginary parts of $V_{ij}$
are independent zero centered Gaussian random variables with variance 
satisfying,
\be
\overline{V_{ab}(k)\, V_{cd}(k)} = V^2_H\,\delta_{ad}\delta_{bc}\;.
\label{strn-eq3}
\ee
Here the 'over-line' indicates ensemble average. From now on we will drop the hat
over $H$ and denote when needed  $H$ by $H(k)$. Let us add that in physical
systems, $k=2$ is of great interest and in some systems such as atomic nuclei 
it is possible to have $k=3$ and even $k=4$ \cite{Zel3,Fu-13}.

The  $U(N)$ algebra that generates the embedding, as shown in \cite{Ko-05},
gives formulas for the lower order moments of the one-point function, the
eigenvalue density $I(E)=\overline{\lan\lan \delta(H-E)\ran\ran}$ and also for
the two-point function in the eigenvalues. In particular,   explicit formulas
are given in \cite{Ko-05,kota}  for $\overline{\lan H^P\ran^m}$, $P=2,4$ and
$\overline{\lan H^P\ran^m\,\lan H^Q \ran^m}$, $P+Q=2,4$. Used here is the $U(N)$
tensorial decomposition of the  $H(k)$ operator giving $\nu=0,1,\ldots,k$
irreducible parts $B^{\nu , \omega_\nu}(k)$ and  then,
\be
H(k) = \dis\sum^{k}_{\nu =0;\omega_\nu \in \nu} 
W_{\nu , \omega_\nu}(k)\; B^{\nu , \omega_\nu}(k) \;.
\label{strn-eq5}
\ee
With the GUE($k$) representation for the $H(k)$ operator, the expansion 
coefficients W's will be independent zero centered Gaussian random variables 
with, by an extension of Eq. (\ref{strn-eq3}),
\be
\overline{W_ {\nu_1 , \omega_{\nu_1}}(k)\;W_ {\nu_2 , \omega_{\nu_2}}(k)}
= V^2_H\;\delta_{\nu_1 , \nu_2} \delta_{\omega_{\nu_1} \omega_{\nu_2}}\;.
\label{strn-eq6}
\ee
For deriving formulas for the various moments, the first step is to apply the
Wigner-Eckart theorem for the matrix elements of $B^{\nu , \omega_\nu} (k)$.
Given the $m$-fermion states $\l.\l|f_m v_i\r.\ran$, we have with respect to the
$U(N)$ algebra, $f_m=\{1^m\}$, the antisymmetric irreducible representation in
Young tableaux notation and $v_i$ are additional labels. Note that $\nu$
introduced above corresponds to the Young tableaux $\{2^\nu 1^{N-2\nu}\}$ and
$\omega_\nu$ are additional labels. Now, Wigner-Eckart theorem gives
\be
\lan f_m v_f \mid B^{\nu , \omega_\nu}(k) \mid f_m v_i\ran = \lan f_m 
\mid\mid B^{\nu}(k) \mid\mid f_m\ran\;C^{\nu , \omega_\nu}_{f_m v_f\,,\;
\overline{f_m} \overline{v_i}} \;.
\label{W-E}
\ee
Here, $\lan --|| -- || --\ran$ is the reduced matrix element and 
$C^{----}_{----}$ is a $U(N)$ Clebsch-Gordan (C-G) coefficient [note that we are
not making a distinction between $U(N)$ and $SU(N)$]. Also, $\l.\l|
\overline{f_m} \overline{v_i}\r.\ran$ represent a $m$ hole  state (see
\cite{Ko-05} for details). In Young tableaux notation $\overline{f_m} =
\{1^{N-m}\}$. Definition of  $B^{\nu , \omega_\nu}(k)$  and the $U(N)$
Wigner-Racah algebra will give,
\be
\barr{l}
\l|\lan f_m \mid\mid B^{\nu}(k) \mid\mid f_m\ran\r|^2 = 
\Lambda^{\nu}(N,m,m-k)\;,
\;\\
\\
\Lambda^{\mu}(N^\pr,m^\pr,r) = \dis\binom{m^\pr-\mu}{r}\,\dis\binom{
N^\pr-m^\pr+r-\mu}{r}\;.
\earr \label{reduced}
\ee
Note that $\Lambda^{\nu}(N,m,k)$ is nothing but, apart from a $N$ and $m$
dependent factor, a $U(N)$ Racah coefficient \cite{Ko-05}. This and the various
properties of the $U(N)$ Wigner and Racah coefficients give two formulas for the
ensemble  average of a product any two $m$ particle matrix elements of $H$,
\be
\barr{l}
\overline{\lan f_m v_1 \mid H(k) \mid f_m v_2\ran\, \lan f_m v_3 \mid H(k) 
\mid f_m v_4\ran} \\
= V^2_H\;\dis\sum_{\nu=0;\omega_\nu}^{k}\Lambda^{\nu}(N,m,m-k)\; C^{\nu , 
\omega_\nu}_{
f_m v_1\,,\;\overline{f_m} \overline{v_2}}\, C^{\nu , \omega_\nu}_{
f_m v_3\,,\;\overline{f_m} \overline{v_4}}\;,
\earr \label{matrix-ele-a}
\ee
and also
\be
\barr{l}
\overline{\lan f_m v_1 \mid H(k) \mid f_m v_2\ran\, \lan f_m v_3 \mid H(k) 
\mid f_m v_4\ran} \\
= V^2_H\;\dis\sum_{\nu=0;\omega_\nu}^{m-k}\Lambda^{\nu}(N,m,k)\; 
C^{\nu , \omega_\nu}_{
f_m v_1\,,\;\overline{f_m} \overline{v_4}}\, C^{\nu , \omega_\nu}_{
f_m v_3\,,\;\overline{f_m} \overline{v_2}}
\earr \label{matrix-ele-b}
\ee
Eq. (\ref{matrix-ele-b}) follows by applying a Racah transform to the
product of the two C-G coefficients appearing in Eq. (\ref{matrix-ele-a}). 
Let us mention two important properties of the $U(N)$ C-G coefficients that 
are quite useful,
\be
\dis\sum_{v_i} C^{\nu , \omega_\nu}_{f_m v_i\,,\,\overline{f_m} 
\overline{v_i}} =\dis\sqrt{\binom{N}{m}}\;\delta_{\nu , 0}\;,\;\;\;
C^{0,0}_{f_m v_i\,,\;\overline{f_m}
\overline{v_j}}={\binom{N}{m}}^{-1/2}\;\delta_{v_i\,,v_j}\;.
\label{sum00}
\ee 
From now on we will use the symbol $f_m$ only in the C-G coefficients, Racah 
coefficients and the  reduced matrix elements. However, for the matrix elements
of an operator we will use $m$ implying totally antisymmetric state for fermions
(symmetric state for bosons).  

Starting with Eq. (\ref{strn-eq5}) and using Eqs. (\ref{strn-eq6}),
(\ref{matrix-ele-b}) and (\ref{sum00}) will immediately give the formula,  
\be
\overline{\lan [H(k)]^2\ran^m} = {\binom{N}{m}}^{-1}\;\dis\sum_{v_i}\,
\overline{\lan m v_i \mid [H(k)]^2 \mid m v_i\ran} = V^2_H \;\Lambda^0(N,m,k)\;.
\label{strn-eq7a}
\ee
Similarly, for $\overline{\lan H^4\ran^m}$ first the ensemble average is
decomposed into 3 terms as, 
\be
\barr{l}
\overline{\lan\lan [H(k)]^4\ran\ran^m} = \dis\sum_{v_i}\,
\overline{\lan m v_i \mid [H(k)]^4 \mid m v_i\ran} \\
\\
= \dis\sum_{v_i, v_j, v_p , v_l} \l[\overline{\lan m v_i \mid H(k) \mid m 
v_j\ran\,\lan m v_j \mid H(k) \mid m v_p\ran}\;\; \overline{\lan m v_p 
\mid H(k) \mid m 
v_l\ran \lan m v_l \mid H(k) \mid m v_i\ran}\r. \\
\\
+ \overline{\lan m v_i \mid H(k) \mid m v_j\ran\,
\lan m v_l \mid H(k) \mid m v_i\ran}\;\; \overline{\lan m v_j \mid H(k) \mid m 
v_p\ran \lan m v_p \mid H(k) \mid m v_l\ran} \\
\\
+ \l. \overline{\lan m v_i \mid H(k) \mid m v_j\ran\,
\lan m v_p \mid H(k) \mid m v_l\ran}\;\; \overline{\lan m v_j \mid H(k) \mid m 
v_p\ran \lan m v_l \mid H(k) \mid m v_i\ran} \r] \;.
\earr \label{strn-eq7ab}
\ee
Note that the trace $\lan\lan H^4\ran\ran^m = \binom{N}{m} \lan H^4\ran^m$.
It is easy to see that the first two terms simplify to give $2[\overline{\lan
H^2\ran^m}]^2$ and the third term is simplified by applying Eq.
(\ref{matrix-ele-a}) to the first ensemble average and Eq. (\ref{matrix-ele-b})
to the second ensemble average. Then, the final result is
\be
\overline{\lan [H(k)]^4\ran^m} = 2 \l[ \overline{\lan H^2\ran^m}\r]^2 
+ V^4_H \;{\binom{N}{m}}^{-1}\; \dis\sum_{\nu=0}^{min(k,m-k)}\,
\Lambda^{\nu}(N,m,k)\, \Lambda^{\nu}(N,m,m-k)\,d(N:\nu)\;;
\label{strn-eq7b}
\ee
\be
d(N:\nu) = {\dis\binom{N}{\nu}}^2 -{\dis\binom{N}{\nu -1}}^2\;.
\label{eq-dnu}
\ee
Also, a by-product of Eqs. (\ref{matrix-ele-b}) and (\ref{sum00}) is
\be
\dis\sum_{v_j}\,\overline{\lan m v_i \mid H(k) \mid m v_j\ran\, 
\lan m v_j \mid H(k) 
\mid m v_k\ran} = \overline{\lan [H(k)]^2\ran^m}\;\delta_{v_i , v_k} 
\label{sum-hh}
\ee
and we will use this in Section 3. Now we will discuss the results for the
moments of the transition strength densities generated by a transition operator
$\co$.

\section{Lower-order moments of transition strength densities:
$H$ EGUE($k$) and $\co$ an independent EGUE($t$)}

For a spinless fermion system, similar to the $H$ operator, we will consider 
the transition operator $\co$ to be a $t$-body and represented by EGUE($t$) in 
$m$-particle spaces. Then, the matrix of $\co$ in $t$-particle spaces will be 
a GUE with the matrix elements $\co_{ab}(t)$ being zero centered independent 
Gaussian variables with the variance satisfying, 
\be
\overline{\co_{ab}(t)\, \co_{cd}(t)} = V^2_\co\,\delta_{ad}\delta_{bc}\;.
\label{strn-eq3a}
\ee
Further, we will assume that the GUE representing $H$ in $k$ particle space and
the GUE($t$) representing $\co$ in $t$ particle space are independent (this is
equivalent to the statement that $\co$ do not generate diagonal elements $\lan
E_i \mid \co \mid E_i\ran$; see \cite{FKPT}). It is important to mention that in
the past there are attempts to study numerically transition strengths using the
eigenvectors generated by a EGOE (with symmetries) for $H$ for nuclear systems
and a realistic transition operator with results in variance with those obtained
from realistic interactions \cite{Zel}. As described in the present paper, a
proper random matrix theory for transition strengths has to employ ensemble
representation for both the Hamiltonian and the transition operator.

With EGUE representation, $\co$ is Hermitian and hence $\cod=\co$. Moments of
the transition strength densities $I_\co(E_i,E_f) $ are defined by
\be
M_{PQ}(m) = \overline{\lan \cod H^Q \co H^P\ran^m} = \overline{\lan \co H^Q 
\co H^P\ran^m}\;.
\label{strn-eq4}
\ee
Here the ensemble average is w.r.t. both EGUE($k$) and EGUE($t$). Now we will
derive formulas for $M_{PQ}$ with $P+Q=2$ and $4$; the moments with $P+Q$ odd
will vanish by definition. 

Firstly, the unitary decomposition of $\co(t)$ gives, 
\be 
\co(t) =
\dis\sum^{t}_{\nu =0;\omega_\nu \in \nu}  U_{\nu , \omega_\nu}(t)\; B^{\nu ,
\omega_\nu}(t) \;. 
\label{strn-eq5a} 
\ee 
The $U$'s satisfy a relation similar to Eq. (\ref{strn-eq6}). Now, for $P=Q=0$
we have using Eq. (\ref{strn-eq7a}), 
\be
\overline{\lan \co \co\ran^m} = V^2_\co \; \Lambda^0(N,m,t) \;. 
\label{strn-eq8}
\ee 
Also, we have the relations 
\be 
\overline{\lan \co \co H^P\ran^m} =
\overline{\lan \co \co\ran^m}\; \overline{\lan H^P\ran^m} = \overline{\lan \co
H^P \co\ran^m}  \;.
\label{strn-eq9} 
\ee 
For the proof consider $\overline{\lan \co \co H^P\ran^m} ={\binom{N}{m}
}^{-1}\; \sum_{v_i,v_j}\, \overline{\lan m v_i \mid \co \co \mid m v_j\ran} \;\;
\overline{\lan m v_j \mid H^P \mid m v_i\ran}$. Now, applying Eq.
(\ref{sum-hh}) gives $\overline{\lan m v_i \mid \co(t) \co(t) \mid m v_j\ran} = 
\overline{\lan \co \co\ran^m}\,\delta_{v_i , v_j}$ and substituting this will
give the first equality in Eq. (\ref{strn-eq9}). Finally, $\overline{\lan \co
\co H^P\ran^m}$ = $\overline{\lan \co H^P \co\ran^m}$ follows from the cyclic
invariance of the $m$-particle average. Eq. (\ref{strn-eq9}) gives, 
\be  
M_{20}(m) = M_{02}(m) = \overline{\lan \co
\co\ran^m} \;\;\overline{ \lan H^2\ran^m}\;,\;\; M_{40}(m) = M_{04}(m) =
\overline{\lan \co \co\ran^m} \;\;\overline{ \lan H^4\ran^m}\;. 
\label{strn-eq9c} 
\ee 
Formulas for $\overline{\lan \co \co\ran^m}$, $\overline{\lan H^2\ran^m}$ and
$\overline{\lan H^4\ran^m}$ follow from Eqs. (\ref{strn-eq8}), (\ref{strn-eq7a})
and (\ref{strn-eq7b}). Thus, the non-trivial moments $M_{PQ}$ for  $P+Q \leq 4$
are $M_{11}$, $M_{13}=M_{31}$ and $M_{22}$.

It is easy to recognize that the bivariate moment $M_{11}$ has same structure as
the third term in Eq. (\ref{strn-eq7ab}) for $\overline{\lan H^4\ran^m}$ as the
$\co$ and $H$ ensembles are independent. Then, the formula for $M_{11}$ follows
directly from the second term in Eq. (\ref{strn-eq7b}). This gives,
\be
\barr{l}
M_{11}(m) =\overline{\lan \co(t)\, H(k)\,\co(t)\,H(k)\ran^m} = \\
V^2_\co\; V^2_H\;{\binom{N}{m}}^{-1}\; \dis\sum_{\nu=0}^{min(k,m-t)}\,
\Lambda^{\nu}(N,m,t)\, 
\Lambda^{\nu}(N,m,m-k)\,d(N:\nu)\;.
\earr \label{strn-eq10}
\ee
This equation has the correct $ t \leftrightarrow k$ symmetry. Using Eqs. (
\ref{strn-eq10}) and (\ref{strn-eq7a}), we have for the bivariate correlation
coefficient $\xi$, the formula
\be
\xi(m) = \dis\frac{M_{11}(m)}{\dis\sqrt{M_{20}(m)\, M_{02}(m)}} = 
\dis\frac{\dis\sum_{\nu=0}^{min(t,m-k)}\,
\Lambda^{\nu}(N,m,k)\, \Lambda^{\nu}(N,m,m-t)\,d(N:\nu)}{
\binom{N}{m}\;\Lambda^0(N,m,t)\,\Lambda^0(N,m,k)}\;.
\label{strn-eq11}
\ee
Turning to $M_{PQ}$ with $P+Q=4$, the first trivial moment is $M_{13}=M_{31}$. 
For $M_{31}$ we have,
\be
\barr{l}
\binom{N}{m}\;M_{31}(m)=\overline{\lan\lan \co(t) H(k) \co(t) [H(k)]^3
\ran\ran^m} \\
= \dis\sum_{v_i, v_j,
v_k, v_l} \overline{\lan m v_i \mid \co(t) \mid m v_j\ran\,\lan m v_k \mid 
\co(t) \mid m v_l\ran}\;\overline{ \lan m v_j \mid H(k) \mid
m v_k\ran\,\lan m v_l \mid [H(k)]^3 \mid m v_i\ran} 
\earr \label{eq-m31a}
\ee
and the last equality follows from the result that the EGUE's representing 
$H$ and $\co$ are independent. The ensemble average of the product of two $\co$
matrix elements follows easily from Eq. (\ref{matrix-ele-b}) giving, 
\be
\overline{\lan m v_i \mid \co(t) \mid m v_j\ran\,\lan m v_k \mid \co(t) \mid m 
v_l\ran} = V^2_\co\dis\sum_{\nu=0}^{m-t} \Lambda^{\nu}(N,m,t) \;\dis\sum_{\omega_\nu}
C^{\nu , \omega_\nu}_{f_m v_i\,,\;\overline{f_m} \overline{v_l}}\, 
C^{\nu , \omega_\nu}_{f_m v_k\,,\;\overline{f_m} \overline{v_j}}\;.
\label{eq-m31b}
\ee
The ensemble average of the product of a $H$ matrix element and $H^3$ matrix
element appearing in Eq. (\ref{eq-m31a}) is,
\be
\barr{l}
\overline{ \lan m v_j \mid H(k) \mid m v_k\ran\,\lan m v_l \mid [H(k)]^3 
\mid m v_i\ran} = \\
\\
\dis\sum_{v_p , v_q} \l[\overline{\lan m v_j \mid H(k) \mid m v_k\ran\,
\lan m v_l \mid H(k) \mid m v_p\ran}\;\; \overline{\lan m v_p \mid H(k) \mid m 
v_q\ran \lan m v_q \mid H(k) \mid m v_i\ran}\r. \\
\\
+ \overline{\lan m v_j \mid H(k) \mid m v_k\ran\,
\lan m v_q \mid H(k) \mid m v_i\ran}\;\; \overline{\lan m v_l \mid H(k) \mid m 
v_p\ran \lan m v_p \mid H(k) \mid m v_q\ran} \\
\\
+ \l. \overline{\lan m v_j \mid H(k) \mid m v_k\ran\,
\lan m v_p \mid H(k) \mid m v_q\ran}\;\; \overline{\lan m v_l \mid H(k) \mid m 
v_p\ran \lan m v_q \mid H(k) \mid m v_i\ran} \r] \;.
\earr \label{eq-m31c}
\ee
The first two terms in Eq. (\ref{eq-m31c}) combined with Eq. (\ref{eq-m31a}) 
will give $2\overline{\lan [H(k)]^2\ran^m}\,M_{11}(m)$. For the third term 
we can use Eqs. (\ref{matrix-ele-a}) and (\ref{matrix-ele-b}) giving
\be
\barr{l}
\overline{\lan m v_j \mid H(k) \mid m v_k\ran\,
\lan m v_p \mid H(k) \mid m v_q\ran}\;\; \overline{\lan m v_l \mid H(k) 
\mid m v_p\ran \lan m v_q \mid H(k) \mid m v_i\ran} \\
\\
= V_H^4 \dis\sum_{\nu_1=0;\omega_{\nu_1}}^{k} 
\dis\sum_{\nu_2=0;\omega_{\nu_2}}^{m-k} \Lambda^{\nu_1}(N,m,m-k)\;
\Lambda^{\nu_2}(N,m,k) \\
\\
\times\; C^{\nu_1 , \omega_{\nu_1}}_{f_m v_j\,,\;\overline{f_m} 
\overline{v_k}}\, 
C^{\nu_1 , \omega_{\nu_1}}_{f_m v_p\,,\;\overline{f_m} \overline{v_q}}\,
C^{\nu_2 , \omega_{\nu_2}}_{f_m v_l\,,\;\overline{f_m} \overline{v_i}}\, 
C^{\nu_2 , \omega_{\nu_2}}_{f_m v_q\,,\;\overline{f_m} \overline{v_p}}\;.
\earr \label{eq-m31d}
\ee
Combining this with Eq. (\ref{eq-m31b}) and applying the orthonormal properties
of the C-G coefficients will give the final formula for $M_{31}$,
\be
\barr{l}
M_{31}(m) =\overline{\lan \co(t)\, H(k)\,\co(t)\,[H(k)]^3\ran^m} = \\
\\
V^2_\co\; V^4_H\,{
\binom{N}{m}}^{-1} \l\{2\Lambda^0(N,m,k)\;\dis\sum_{\nu=0}^{min(k,m-t)}\,
\Lambda^{\nu}(N,m,t)\, \Lambda^{\nu}(N,m,m-k)\,d(N:\nu) \r. \\
+ \l. \dis\sum_{\nu=0}^{min(k,m-k,m-t)}\,\Lambda^{\nu}(N,m,t)\, 
\Lambda^{\nu}(N,m,k)\,\Lambda^{\nu}(N,m,m-k)\,d(N:\nu)\r\}\;. 
\earr \label{strn-eq12}
\ee
For deriving the formula for $M_{22}$, we will make use of the decompositions
similar to those in Eqs. (\ref{eq-m31a}) and (\ref{eq-m31c}). Then it is easy to
see that $M_{22}$ will have three terms and simplifying these will give (details
are not given here due to lack of space),
\be
\barr{l}
M_{22}(m) = \overline{\lan O(t)O(t)\ran^m}\;\; \l\{\;\overline{\lan H(k)H(k)
\ran^m}\;\r\}^2 \\
+ V^2_\co V^4_H \,\l\{{\binom{N}{m}}^{-1} \;\dis\sum_{\nu=0}^{
min(t,m-k)}\,\Lambda^{\nu}(N,m,m-t)\, 
\l[\Lambda^{\nu}(N,m,k)\r]^2\,d(N:\nu)\r.  \\
+ \l. {\binom{N}{m}}^{-1}\; \dis\sum_{\nu=0}^{min(k+t,m-k)}
\Lambda^{\nu}(N,m,k)\,d(N:\nu)\;\;\dis\sum_{\nu_1=0}^{t}\; 
\dis\sum_{\nu_2=0}^{k}\; \dis\sum_\rho 
\l|\lan f_m \mid\mid \l[B^{\nu_1}(t) B^{\nu_2}(k)\r]^{\nu\,,\,\rho} \mid\mid 
f_m\ran\r|^2\r\} 
\earr \label{strn-eq13a}
\ee
Here $\rho$ labels multiplicity of the irrep $\nu$ in the Kronecker product
$\nu_1 \times \nu_2 \rightarrow \nu$. The third term above simplifies to,
$$
\barr{l}
{\binom{N}{m}}^{-2} \dis\sum_{\nu=0}^{min(k+t,m-k)}
\;\dis\sum_{\nu_1=0}^{t}\;\; \dis\sum_{\nu_2=0}^{k}\,\Lambda^{\nu}(N,m,k)\, 
\Lambda^{\nu_1}(N,m,m-t)\,\Lambda^{\nu_2}(N,m,m-k) \\
\times d(N:\nu_1) d(N:\nu_2) 
\dis\sum_\rho \l[U(f_m\, \nu_1\,f_m\,\nu_2\;;\;f_m\,\nu)_\rho\r]^2 \;. 
\earr
$$
The $U$ or Racah coefficient here is with respect to $U(N)$.
Deriving formulas for this  $U$-coefficient is an important open problem. 

\section{Asymptotic results for the bivariate moments for $H$ EGUE($k$) and 
$\co$ an independent EGUE($t$)}

Lowest order (sufficient for most purposes) shape parameters of the bivariate
strength density are the bivariate reduced cumulants of order four, i.e.
$k_{rs}$, $r+s=4$. The $k_{rs}$ can be written in terms  of the normalized
central moments  $\wtM_{PQ}$ where $\wtM_{PQ}=M_{PQ}/M_{00}$. Then, the scaled
moments  $\mu_{PQ}$ are 
\be
\mu_{PQ} = \l\{\l[\wtM_{20}\r]^{P/2} 
\l[\wtM_{02}\r]^{Q/2}\r\}^{-1}\;\wtM_{PQ}\;,\;\;\; P+Q \geq 2\;.
\label{eq.normom}
\ee
Note that $\mu_{20}=\mu_{02}=1$ and $\mu_{11}=\xi$. Now the fourth order 
cumulants are,
\be
k_{40} = \mu_{40} - 3\,, \;\;k_{04} = \mu_{04} - 3\,,\;\;
k_{31} = \mu_{31} - 3\,\xi\,,\;\;k_{13} = \mu_{13} - 3\,\xi\,,\;\;
k_{22}= \mu_{22}- 2\;\xi^2 -1 \;. 
\label{eq.cumu}
\ee
The $\l|k_{rs}\r| \sim 0$ for $r+s \geq 3$ implies that the transition strength
density is close to a bivariate Gaussian (note that in our EGUE applications,
$k_{rs}=0$ for $r+s$ odd by definition). We have verified in large number of
numerical examples, obtained using the formulas in Section 3 for some typical
values of $N$, $m$, $k$ and $t$, that the cumulants $|k_{rs}|$ with $r+s=4$ are
in general very small implying that for EGUE, transition strength densities
approach bivariate Gaussian form. For a better understanding of this result,  it
is useful to derive expressions for $\mu_{PQ}$ and thereby for $k_{PQ}$, using
Eq. (\ref{eq.cumu}), in the asymptotic limit defined by  $N \rightarrow \infty$,
$m \rightarrow \infty$, $m/N \rightarrow 0$  and $k$ and $t$ fixed. First we 
will consider $N \rightarrow \infty$ and $m$ fixed with $k,t << m$. We will use,
\be 
\binom{N-p}{r} \stackrel{p/N \rightarrow 0}{\longrightarrow} 
\dis\frac{N^r}{r!}\;,\;\; d(N:\nu) \stackrel{\nu/N \rightarrow
0}{\longrightarrow} \dis\frac{N^{2\nu}}{ (\nu!)^2}\;. 
\label{asymp2} 
\ee 
Note that $d(N:\nu)$ was defined by Eq. (\ref{eq-dnu}). Let us start with the
formula for $\xi$ given  by Eq. (\ref{strn-eq11}). Applying Eq. (\ref{asymp2})
it is easily seen that only the term with $\nu=t$ in the sum in Eq.
(\ref{strn-eq11}) will contribute in the asymptotic limit. Using this, applying
Eqs. (\ref{asymp2}) and the formula for $\Lambda^\nu$ given by Eq. 
(\ref{reduced}) will lead to, 
\be 
\xi(m)  \longrightarrow  \dis\frac{m!\;N^{2t}}{N^m\;(t!)^2} 
\dis\frac{\binom{m-t}{k} \binom{N-m+k-t}{k}\binom{N-2t}{m-t}}{\binom{m}{k} 
\binom{N-m+k}{k} \binom{m}{t}\binom{N-m+t}{t}}  \longrightarrow 
{\binom{m}{k}}^{-1} \binom{m-t}{k}\;. 
\label{asymp3} 
\ee 
Similarly, $\mu_{40}$ will be, using Eqs. (\ref{strn-eq9c}), (\ref{strn-eq7b})
and (\ref{strn-eq7a}),
\be 
\mu_{40}(m)  \longrightarrow  2 + \dis\frac{m!\;N^{2k}}{N^m\;(k!)^2} 
\dis\frac{\binom{m-k}{k} \binom{N-m}{k}\binom{N-2k}{m-k}}{\binom{m}{k} 
\binom{N-m+k}{k} \binom{m}{k}\binom{N-m+k}{k}}  \longrightarrow  2+
{\binom{m}{k}}^{-1} \binom{m-k}{k}\;. 
\label{asymp4} 
\ee 
Turning to $\mu_{31}$, it is easy to see from Eq. (\ref{strn-eq12}) that 
$M_{31}$ has two terms. The first term is $(2\xi)\lan O(t)O(t)\ran^m [\lan H(k)
H(k)\ran^m]^2$ and in the sum in the second term   only $\nu=k$ will survive in
the asymptotic limit. These then will give, 
\be 
\barr{rcl} 
\mu_{31}(m) & = & (2\xi)
+ {\binom{N}{m}}^{-1} \dis\frac{\Lambda^k(N,m,t)  \Lambda^k(N,m,k)
\Lambda^k(N,m,m-k)\,d(N:k)}{\Lambda^0(N,m,t)[ \Lambda^0(N,m,k)]^2} \\ &
\longrightarrow & \xi\,\l[2 + {\binom{m}{k}}^{-1}\,\binom{m-k}{k} \r] = \xi
\mu_{40}\;. 
\earr 
\label{asymp5} 
\ee 
Note that in the final simplifications we have used Eq. (\ref{asymp2}). Finally,
let us consider $\mu_{22}$.  Firstly, $M_{22}$ has three terms as seen from Eq.
(\ref{strn-eq13a}) and then, there will be  corresponding three terms in
$\mu_{22}$. Let us call them $T1$, $T2$ and $T3$.  It is seen that $T1=1$ and
$T2$ is (in the corresponding sum in $M_{22}$, only $\nu=t$ term will  contribute
in the asymptotic limit), 
\be 
T2 \longrightarrow  {\binom{N}{m}}^{-1} \dis\frac{\Lambda^t(N,m,m-t) 
\l[\Lambda^t(N,m,k)\r]^2\,d(N:t)}{\Lambda^0(N,m,t)[\Lambda^0(N,m,k)]^2} 
\longrightarrow  {\binom{m}{k}}^{-2}\,{\binom{m-t}{k}}^2\;. 
\label{asymp6} 
\ee
Similarly, $T_3$ in the asymptotic limit will be 
\be 
T_3 \longrightarrow
\dis\frac{\Lambda^{t+k}(N,m,k)}{\Lambda^{0}(N,m,k)}\; \l[\xi(N \rightarrow
\infty)\r]  \rightarrow {\binom{m}{k}}^{-2} \,\binom{m-t-k}{k} \binom{m-t}{k}
\;. 
\label{asymp9} 
\ee 
Here we have used Eq. (\ref{asymp3}) for $\xi$ in the $N \rightarrow \infty$ 
limit. Now combining $T_1$, $T_2$ and $T_3$ we have 
\be
\mu_{22}(m) \longrightarrow 1 + 
{\binom{m}{k}}^{-2}\,{\binom{m-t}{k}}^2 + {\binom{m}{k}}^{-2}\, \binom{m-t-k}{k}
\binom{m-t}{k}\;. 
\label{asymp10} 
\ee  
Comparing Eq. (\ref{asymp9}) with the formula for $T_3$, it is easy to see  that
$[U(f_m\; t\; f_m\; k\;;\; f_m\; t+k)]^2 \stackrel{asymp}{\longrightarrow} 
{\binom{m}{k}}^{-1}\;\binom{m-t}{k}$.  The asymptotic formulas given by Eqs.
(\ref{asymp3}), (\ref{asymp4}),  (\ref{asymp5}) and (\ref{asymp10}) show that
the finite $N$ results derived in Section 3 reduce in the asymptotic limit
exactly to those derived before using binary correlation approximation
\cite{FKPT,kota}. More importantly, they show that the cumulants tend to 
zero in the dilute (or asymptotic) limit. 

\section{Lower-order moments of transition strength densities:
results for particle transfer operator and beta decay type operator}

In this Section we will consider transition operators that are non-hermitian and
that are of great interest for example in nuclear physics. These are particle
removal (or addition) operators and beta decay type operators.  Let us begin
with particle removal operator $\co$ and say it removes $k_0$ number of
particles when acting on a $m$ fermion state. Then the general form  of $\co$
is,
\be
\co = \dis\sum_{\alpha_0} V_{\alpha_0}\;A_{\alpha_0}(k_0) \;.
\label{snt-1}
\ee
Here, $A_{\alpha_0}(k_0)$ is a $k_0$ particle annihilation operator and
$\alpha_0$ are indices for a $k_0$ particle state. We will choose $V_{\alpha_0}$
to be zero centered independent Gaussian random  variables with variance
satisfying $\overline{V_{\alpha} V^{\dagger}_\beta} =  V^2_\co\;\delta_{\alpha
\beta}$. Then we have,
\be
\overline{\lan \cod \co\ran^m} = V^2_\co\;\binom{m}{k_0}\;,\;\;\;
\overline{\lan \co \cod\ran^m} = V^2_\co\;\binom{N-m}{k_0}\;.
\label{snt-3}
\ee
Eq. (\ref{sum-hh}) also gives the important relations,
\be
\overline{\lan \cod \co H^p\ran^m} = \overline{\lan \cod \co\ran^m}\;\; 
\overline{\lan H^p\ran^m}\;,\;\;\;
\overline{\lan \cod H^p \co\ran^m} = \overline{\lan \cod \co\ran^m}
\;\; \overline{\lan H^p\ran^{m-k_0}}\;.
\label{snt-4}
\ee
Another useful result that follows from Eq. (\ref{snt-3}) is,
\be
\lan m \mid\mid A^{\dagger}(k_0) \mid\mid m-k_0\ran \lan m-k_0 \mid\mid
A(k_0) \mid\mid m\ran = \binom{N-k_0}{m-k_0} \;.
\label{snt4a}
\ee
Following the procedure used in Section 3 it is possible to derive formulas for
the lower order moments of the transition strength density generated by $\co$ 
given by Eq. (\ref{snt-1}). Eqs. (\ref{snt-4}) and (\ref{snt-3}) along with Eqs.
(\ref{strn-eq7a}) and (\ref{strn-eq7b}) will give $M_{20}$, $M_{02}$, $M_{40}$ 
and $M_{04}$. Then, the first non-trivial moment is $M_{11}$. Formula for 
$M_{11}=\overline{\lan \cod H \co H\ran^m}$ is derived by starting with the
definition of $\overline{\lan \cod H \co H\ran^m}$,
\be
\barr{l}
\binom{N}{m} \; \overline{\lan \cod H \co H\ran^m} = 
\dis\sum_{v_1, v_2, v_3, v_4} \overline{
\lan m, v_1 \mid \cod \mid m-k_0, v_2\ran \lan m-k_0, v_3 \mid \co \mid 
m, v_4\ran}\\
\times\;\overline{\lan m-k_0, v_2 \mid H \mid m-k_0, v_3\ran \lan 
m, v_4 \mid H \mid m, v_1\ran}\;. 
\earr \label{snt-5}
\ee
Applying Eqs. (\ref{strn-eq5}) - (\ref{reduced}) and the Wigner-Eckart theorem 
along with Eq. (\ref{snt4a}) will give,
\be
\barr{l}
M_{11}(m) = \overline{\lan \cod H(k) \co H(k)\ran^m} =  V_{\co}^2 V_H^2\;
{\binom{N}{m}}^{-1}\; \binom{N-k_0}{m-k_0}\; {\binom{N}{k_0}}^{1/2}
\times \dis\sum_{\nu=0}^k \\
\l[d(N:\nu)\,\Lambda^{\nu}(N,m-k_0,m-k_0-k)\;
\Lambda^{\nu}
(N,m,m-k)\r]^{1/2}\;(-1)^{\phi}\;U(f_{m}\,\overline{f_{m-k_0}}\,f_{m}\,
f_{m-k_0}\,;\,f_{k_0}\,\nu)\;.
\earr \label{snt-7}
\ee
The phase factor $\phi$ depends on $\nu$. Formula for the $U$-coefficient
appearing in Eq. (\ref{snt-7}) was available in \cite{He-75}. It is possible to
proceed forward and derive formulas for $M_{31}$, $M_{13}$ and $M_{22}$ (these
will be reported elsewhere). Now we will turn to beta decay type operators.

Let us consider a system with $m_1$ fermions in $N_1$ sp states and $m_2$
fermions in $N_2$ sp states with $H$ preserving $(m_1,m_2)$. Then, the $H$ 
operator, assumed to be $k$-body, is given by,
\be
\barr{l}
H(k) = \dis\sum_{i+j=k} \dis\sum_{\alpha , \beta \in i} \dis\sum_{a,b \in j}\;
V_{\alpha a:\beta b}(i,j)\;
A^\dagger_{\alpha}(i)\, A_{\beta}(i)\, A^\dagger_a(j)\,A_b(j)\;,\\
V_{\alpha a:\beta b}(i,j) = \lan i, \alpha : j, a \mid H \mid i, \beta : 
j,b\ran\;.
\earr \label{strn-eq14}
\ee
Here we are using Greek labels $\alpha, \beta, \ldots$ to denote the many
particle states generated by fermions occupying the orbit with $N_1$ sp states
and the Roman labels $a,b,\ldots$ for the many particle states generated by the
fermions occupying the orbit with $N_2$ sp states.  For each $(i,j)$ pair with
$i+j=k$, we have a matrix $V(i,j)$ in the $k$-particle space and we assume that
the $V(i,j)$ matrices are represented by independent GUE's with 
their matrix elements being zero centered with variance,
\be
\overline{V_{\alpha a : \beta b}(i,j)\;V_{\alpha^\pr a^\pr : \beta^\pr b^\pr}
(i^\pr ,j^\pr)} 
=V^2_H(i,j)\;\delta_{i i^\pr} \delta_{j j^\pr} \delta_{\alpha \beta^\pr} 
\delta_{a b^\pr} \delta_{\beta \alpha^\pr} \delta_{b a^\pr} \;.
\label{strn-eq15}
\ee
It is important to note that the embedding algebra for the EGUE generated by 
the action of the $H(k)$ operator on $\l.\l|m_1, v_\alpha : m_2, v_a\r.\ran$ 
states is the direct sum algebra  $U(N_1) \oplus U(N_2)$. Thus we have 
EGUE($k$)-$[U(N_1) \oplus U(N_2)]$ ensemble. For this system,  we will 
consider a beta decay ($k_0=1$ below) type transition operator,
\be
\co = \dis\sum_{\alpha , a} O_{\alpha a} A^\dagger_{\alpha}(k_0)\,
A_a(k_0)\;;\;\;\;O_{\alpha a} =
\lan k_0,\alpha \mid \co \mid k_0, a\ran
\label{strn-eq16}
\ee
and assume a GUE representation for the $\co$ matrix in the defining space
giving $\overline{O^\dagger_{\alpha , a} O_{\beta , b}}=
V^2_\co\;\delta_{\alpha \beta}  \delta_{a b}$.  Note that in general the $\co$
matrix is a rectangular matrix. Now,  the ensemble averaged bivariate moments of
the transition strength density are $M_{PQ}(m_1 , m_2)=\overline{\lan \cod H^Q
\co H^P\ran^{(m_1 , m_2)}}$. Note that $\co$ takes $(m_1 , m_2)$ to $(m_1+k_0 ,
m_2-k_0)$. 

Firstly, recognize that $A^\dagger_{\alpha}(k_0)$ and $A_\beta(k_0)$ transform
as the $U(N_1)$ tensors $f_{k_0}=\{1^{k_0}\}$ and $\{\overline{f_{k_0}}\}=
\{1^{N_1-k_0}\}$ and similarly $A^\dagger_{a}(k_0)$ and $A_b(k_0)$ with respect
to $U(N_2)$. These will give easily the results,
\be
\overline{\lan \cod \co\ran^{(m_1 , m_2)}} = V^2_\co\;\binom{N_1 -m_1}{k_0}\,
\binom{m_2}{k_0}\;,\;\;\;
\overline{\lan \co \cod \ran^{(m_1 , m_2)}} = V^2_\co\;\binom{N_2 -m_2}{k_0}\,
\binom{m_1}{k_0}\;.
\label{strn-eq20}
\ee
For deriving formulas for the bivariate moments with $P$ and/or $Q \neq 0$, 
unitary decomposition of $H$ is carried out with respect to the $U(N_1) \oplus 
U(N_2)$ algebra. This gives, 
\be
\lan H^2\ran^{(m_1 , m_2)} = \dis\sum_{i+j=k}\,V^2_H(i,j)\;\Lambda^0(N_1, m_1, 
i) \,\Lambda^0(N_2, m_2, j)\;.
\label{strn-eq22}
\ee
It is also easy to show that $\overline{\lan \cod \co H^P\ran^{(m_1,m_2)}} =
\overline{\lan \cod \co  \ran^{(m_1,m_2)}}\; \overline{\lan H^P
\ran^{(m_1,m_2)}}$ and $\overline{\lan \cod H^P \co\ran^{(m_1,m_2)}} =
\overline{\lan \cod \co  \ran^{(m_1,m_2)}}\; \overline{\lan H^P \ran^{(m_1+k_0,
m_2-k_0)}}$. Also, the first nontrivial bivariate moment $M_{11}(m_1 , m_2) =
\overline{\lan \cod H \co H\ran^{(m_1,m_2)}}$ is given by,
\be
\barr{l}
M_{11}(m_1,m_2) =
V^2_\co\;\binom{N_1-k_0}{m_1}\;\binom{N_2-k_0}{m_2-k_0}\;\l[\binom{N_1}{m_1}\;
\binom{N_2}{m_2}\r]^{-1}\;
\dis\sqrt{\binom{N_1}{k_0}\;\binom{N_2}{k_0}}\;\;\dis\sum_{i+j=k} V^2_H(i,j) \\
\times \;\dis\sum_{\nu=0}^{i}\;\; 
\dis\sum_{\nu^\pr = 0}^{j}\;\l[d(N_1 : \nu)\;d(N_2 : \nu^\pr)\,
\Lambda^{\nu}(N_1, m_1, m_1-i)\Lambda^{\nu}(N_1, m_1+k_0, 
m_1+k_0-i)\r]^{1/2} \\
\times \;\l[\Lambda^{\nu^\pr}(N_2, m_2, m_2-j)\Lambda^{\nu^\pr}(N_2, m_2-k_0, 
m_2-k_0-j)\r]^{1/2} \;\;(-1)^{\phi_1(\nu) + \phi_2(\nu^\pr)}\\
\times \;\;U(f_{m_1+k_0}\,\overline{f_{m_1}}\,f_{m_1+k_0}\,f_{m_1}\,;\,
f_{k_0}\,\nu)\;  
U(f_{m_2}\,\overline{f_{m_2-k_0}}\,f_{m_2}\,f_{m_2-k_0}\,;\,f_{k_0}\,\nu^\pr)\;.
\earr \label{strn-eq25}
\ee
Here, $\phi_1$ and $\phi_2$ are phase factors. Also, $\overline{f_r} = \{1^{N-r
}\}$ and $\nu=\{2^\nu , 1^{N-2\nu}\}$ with $N=N_1$ or $N_2$ as appropriate. 
Formula for the $U$-coefficients in Eq. (\ref{strn-eq25}) is available in
\cite{He-75}.

\section{Conclusions and future outlook}

In this paper we have presented for the first time exact (finite $N$) results
for the moments of the transition strength densities using $U(N)$ Wigner-Racah
algebra for EGUE random matrix ensembles. In particular, formulas for the
moments up to fourth order are derived in detail for the Hamiltonian a EGUE($k)$
and the transition operator a EGUE($t$) for spinless fermion systems. Numerical
results on one hand and the asymptotic results derived from the exact results on
the other, showed that the fourth order cumulants approach zero in the dilute
limit implying that the strength densities approach bivariate Gaussian form. As
discussed in Section 5, the formulation given in Sections 2 and 3 extends to
transition operators  that are particle removal (or particle addition) operators
and also to beta decay (also neutrinoless double beta decay) type operators.
Complete results for these (i.e. for $M_{PQ}$ with $P+Q=4$) will be presented
elsewhere. All these results are useful in nuclear spectroscopy
\cite{FKPT,kota}. Another important extension is to EGUE with $U(\Omega) \times
SU(r)$ embedding discussed in \cite{Ma-12}. With applications in mesoscopic
systems, it is important to derive formulas for the bivariate moments of
transition strength densities for systems with $r=2$. For fermions, this
corresponds to spin degree of freedom and then we have spin scalar and spin
vector transition operators. For this system, formulas for the  Racah
coefficients that appear, even for the fourth moment of the state densities, are
not yet available \cite{kota} and an exploration of the asymptotic methods
suggested in \cite{Fr-79} could prove to be fruitful. Let us mention that the
results presented in Sections 3 and 5 extend to boson systems by using the $N
\rightarrow -N$ and $N \rightarrow m$ symmetries discussed in \cite{Ko-05,kota}.
In future, it is also important and useful to extend the present work to EGOE
and EGSE ensembles. Finally, future  in the subject of embedded random matrix
ensembles in quantum physics is exciting with enormous scope for developing new
group theoretical methods for their analysis and with the possibility of their
applications in a variety of isolated finite quantum many-particle systems.

\ack
Thanks are due to Prof. H.A. Weidenm\"{u}ller, for many useful discussions.

\end{document}